\documentclass[]{aa}

\usepackage{graphicx}
\usepackage{amsmath,textcomp}
\usepackage{txfonts}
\usepackage{hyperref}

\def\vect#1{\boldsymbol{#1}}
\newcommand{\mathsbf}[1]{{\boldsymbol{\mathsf{#1}}}}

\usepackage{ulem}
\usepackage{xcolor}

\begin{document} 

\title{Statistical properties of compressible isothermal turbulence from sub- to supersonic conditions.}

\titlerunning{Compressible isothermal turbulence}

\author{F. Thiesset\inst{\ref{inst1}}
\and 
C. Federrath\inst{\ref{inst2}}$^,$\inst{\ref{inst3}}} 
\institute{CNRS, CORIA, UMR 6614, Normandy Univ., UNIROUEN, INSA Rouen, 675 Avenue de l'université, BP 12, 76801 Saint Etienne du Rouvray Cedex
\email{fabien.thiesset@cnrs.fr} \label{inst1}
\and
Research School of Astronomy and Astrophysics, Australian National University, Cotter Road, Canberra, ACT 2611, Australia \label{inst2}
\and 
Australian Research Council Centre of Excellence in All Sky Astrophysics (ASTRO3D), Cotter Road, Canberra, ACT 2611, Australia  \\
\email{christoph.federrath@anu.edu.au}\label{inst3} 
}

\authorrunning{F. Thiesset \and C. Federrath}

\abstract
    {Turbulence is one of the key processes that control the spatial and temporal evolution of matter and energy of many astrophysical systems.}
    {This paper investigates the statistical properties of isothermal turbulence in both the subsonic and supersonic regimes. The focus is on the influence of the Mach number ($Ma$) and the Reynolds number ($Re$) on both the space-local and scale-dependent fluctuations of relevant gas variables, the density, velocity, their derivatives, and the kinetic energy.}
    {We carry out hydrodynamical simulations of driven turbulence with explicit viscosity and therefore controlled $Re$, at converged numerical resolutions up to $1920^3$~grid cells.}
    {We confirm previous work that the probability density functions (PDFs) of the gas density are approximately log-normal and depend on $Ma$. We provide a new detailed quantification of the dependence of the PDFs of density and velocity on $Re$, finding a relatively weak dependence, provided $Re>200$. In contrast, derivatives of the density and velocity field are sensitive to $Re$, with the probability of extreme events (the tails of the PDFs) growing with $Re$. The PDFs of the density gradient and velocity divergence (dilatation) exhibit increasingly heavy tails with growing $Ma$, signalling enhanced internal intermittency. At sufficiently high $Ma$, the statistics of dilatation are observed to saturate at a level determined solely by $Re$, suggesting that turbulent dilatation becomes limited by viscous effects. We also examine the scale-by-scale distribution of kinetic energy through a compressible form of the Kármán-Howarth-Monin (KHM) equation that incorporates density variations. In the intermediate range of scales, a marked difference is found between subsonic and supersonic turbulence: while Kolmogorov-like scaling applies in the sub- and transonic regimes, supersonic turbulence aligns more closely with Burgers turbulence predictions. The analysis of individual terms in the KHM equation highlights the role of the pressure-velocity coupling as an additional mechanism for converting kinetic energy from large to small scales. Moreover, the contributions of the KHM terms exhibit non-monotonic behaviour with increasing $Ma$, with dilatational effects becoming more pronounced and acting to oppose the cascade of kinetic energy.}
    {}


\maketitle

\section{Introduction}

Turbulence reigns as the most ubiquitous and yet most enigmatic face of fluid movement. It drives the transport of matter at nearly all scales from micro- to astrophysical scales \citep{Dubrulle2019}. In situations where the characteristic speed of a fluid is much less than the speed of sound in the surrounding medium, the assumption of incompressible flow offers a useful simplification. Most current knowledge pertains to this category of fluid flows. In this context, turbulence is recognised as a paradigmatic out-of-equilibrium system where the energy injected at large scales is transferred across intermediate scales via a nonlinear cascade process, before being dissipated at the smallest scales. The characterisation and prediction of velocity fluctuations at different flow scales constitutes a fundamental pursuit in the field of fluid mechanics.

In astrophysical contexts, fluid flows are however best described as compressible and/or with significant density fluctuations, a characteristic that also emerges across structures of vastly different cosmological scales, from the intergalactic medium, the interstellar medium, to accretion discs and stellar interiors. Owing to their immense scale and mass, it is then more accurate to recognise compressible and/or variable-density turbulence as the prevailing state of fluid motion. In the interstellar medium {(ISM)}, and particularly within cold, dense molecular clouds that serve as stellar nurseries, compressible turbulence plays an ambivalent role \cite[see for example refs.][and references therein]{Elmegreen2004,MacLow2004,McKee2007,Hennebelle2012}. It contributes to the formation of dense clumps that triggers gravitational collapse, while simultaneously injecting turbulence that counteracts this collapse and inhibits star formation. A detailed understanding of compressible turbulence is thus crucial for the characterisation of the structure and dynamics of most astrophysical systems.

Compressible turbulence is distinguished from its incompressible counterpart by the interdependence of fluid motion and thermodynamic properties, manifesting as coupled fluctuations in velocity, pressure, and density \cite[see e.g ref.][and references therein]{Rabatin2023,Scannapieco2024}. Then, the thermodynamics of the fluid matters \citep{Passot1998,Banerjee2014,Sakurai2024}. This coupling also yields energy conversion processes, particularly between the kinetic energy and internal energy reservoirs, thereby enriching the complexity of the mechanisms at play \citep{Galtier2011,Aluie2012,Wang2018,Schmidt2019}. For astrophysical applications, additional forms of energy such as magnetic \citep[e.g.][]{Banerjee2013,Seta2021} and gravitational potential energy \citep[e.g.][]{Banerjee2018} must also be considered, further complicating the picture of energy exchange processes.

{Observations indicate that the interstellar medium (ISM) is highly inhomogeneous \citep{Myers1978}, with regions of low and high turbulence intensity often coexisting in close proximity \citep{Linsky2022}. As a result, the local turbulent conditions within a sub-region of the ISM may differ significantly from those inferred from cloud-averaged properties. Rather than introducing further complexity by explicitly modelling this inhomogeneity, we adopt a simplified approach: we treat the ISM as a collection of locally homogeneous sub-regions, each characterised by distinct turbulent properties. Our aim is to systematically characterise the statistical signatures associated with different turbulent regimes. We focus in particular on the statistics of velocity and density. For simplicity, we also assume the ISM to be isothermal, an assumption generally justified in dense molecular clouds, where radiative cooling is highly efficient \citep{Elmegreen2004,Glover2010,Krumholz2014}. Despite its simplicity, this framework offers a useful first step toward capturing the statistical diversity of ISM substructure.}

In practice, homogeneous turbulence can be readily realised through direct numerical simulations (DNS) of the compressible Navier-Stokes equation \citep{Pirozzoli2004,Donzis2013,Jagannathan2016,John2021,Sakurai2021,Sakurai2023,Sakurai2024,Wang2011,Wang2017,Wang2018,Seta2021} or implicit large eddy simulation (ILES) of the Euler equation in a periodic box \citep{Kritsuk2007,Federrath2008,Schmidt2009,Federrath2010,Aluie2012,Pan2019a,Federrath2021,Rabatin2023,Scannapieco2024,Beattie2025}. DNS of the Navier-Stokes equation can now achieve relatively large $Re$ although generally restricted to relatively low Mach, from the sub- to transonic regimes. In contrast, ILES of the Euler-equation are typically used to explore the supersonic regime but with no explicit control of the viscous cut-off.

A substantial body of work has now established a relatively coherent picture of the statistical properties of compressible turbulence. For instance, the Mach-number dependence of fluctuations of the density, velocity, and pressure, and their spatial derivatives are well documented \citep{Passot1998,Pirozzoli2004,Kritsuk2007,Federrath2008,Schmidt2009,Federrath2010,Wang2011,Konstandin2012,Donzis2013,Wang2017,Pan2019,Rabatin2023}. {However, there are still unexplored regions of the parameter space; in particular, the $Re$ dependence of velocity and density statistics is largely unknown as most previous work in the supersonic regime are ILES.} 

{We believe that examining the Reynolds number dependence of velocity and density statistics is important in an astrophysical context. First, in the ISM, where turbulent activity varies across subregions, the dependence of certain statistics on Reynolds numbers can reveal key information about its local structure. Second, many processes including heating and chemical reactions in the ISM, are sensitive to small-scale turbulence, particularly to kinetic energy dissipation \citep{Godard2009}. Direct Numerical Simulations (DNS) with explicitly prescribed Reynolds numbers thus offer valuable insight into small-scale dynamics and help constrain models of the ISM’s thermal and chemical energy budgets. This could not be achieved with ILES, because they lack control over the viscous cutoff scale, leading to erroneous statistics of the velocity dilatation and kinetic energy dissipation rate \citep{Pan2019,Pan2019a}. Third, the ISM exhibits internal intermittency \citep{Falgarone1995,Falgarone2006,HilyBlant2007,HilyBlant2008,Falgarone2009}, manifesting as extreme, localised fluctuations in gradients. For compressible, possibly supersonic turbulence, the dependence of such extreme events on Reynolds number remains an open question, which this study aims to clarify. Finally, while the ISM typically operates at Reynolds numbers exceeding those accessible in simulation, it remains unclear at what Reynolds number and at which rate, turbulence statistics converge to their asymptotic behaviour, if one were to exist \citep{John2021}. Exploring this transition in the supersonic regime helps clarify the limitations of current simulations when compared to observations.}

{As the ISM is a highly inhomogeneous medium, it is essential to investigate not only the spatial fluctuations of field variables but also their dependence on scale. Probability density functions (PDFs) capture spatial variability, while additional statistical tools are required to characterise the flow properties across different scales. Together, these tools offer a detailed view of ISM sub-structure.} A variety of scale-dependent analytical approaches has been developed, which include analysis in Fourier space \citep{Kritsuk2007,Federrath2010,Wang2013,Wang2017a,Schmidt2019,Hellinger2021}, coarse-graining \citep{Aluie2012,Aluie2013,Wang2013,Wang2018}, and point-splitting methods (correlation or structure functions) \citep{Kritsuk2007,Federrath2009,Falkovich2010,Galtier2011,Konstandin2012,Wang2013,Banerjee2013,Banerjee2014,Banerjee2018,Ferrand2020,Hellinger2021,Pan2022}. {In astrophysical contexts, because the scale at which e.g. a star is forming can differ from the scale at which turbulence is driven \citep{Elmegreen2008,Federrath2017}, understanding how and at which scale energy is injected, transferred, converted and dissipated is crucial. This information is again valuable for developing a more representative picture of the ISM sub-structure and ultimately more accurate models for the star formation rate \citep{Padoan2011,Hennebelle2011,Federrath2012}.} The influence of $Ma$ and $Re$ on the scale-dependent processes was characterised in some previous work \citep{Schmidt2019,Hellinger2021,Wang2018}. Here we significantly extend these works by considering a larger range of $Ma$ and $Re$. Of particular interest is also the convergence of scale-local mechanisms at asymptotically large $Re$--particularly the development of an inertial range \citep{Aluie2013}--mirroring trends observed in high-$Re$ incompressible turbulence \citep{Antonia2006,Danaila2012}. An analogous question arises in compressible flows in the limit of infinitely large $Ma$.

In the present study, we perform a statistical analysis of isothermal, statistically stationary, homogeneous, and isotropic compressible turbulence, using high-fidelity numerical simulations, spanning a wide range of $Ma$, from subsonic to supersonic regimes, and controlled $Re$ through DNS. We begin by examining the space-local quantities by quantifying the PDFs of the density and velocity, along with their spatial derivatives. Subsequently, we analyse the scale-dependence of the kinetic energy using a point-splitting approach based on a compressible extension of the Kármán-Howarth-Monin equation.

The remainder of the paper is organised as follows. The numerical simulations are described in section~\ref{sec:dns}. Section~\ref{sec:results_pdf} presents the results concerning the PDFs of the velocity and density and their derivatives, while Section~\ref{sec:results_sbs} is dedicated to the scale-by-scale energy analysis. Finally, the main conclusions are summarised in Section~\ref{sec:summary}.

\section{Numerical simulations} \label{sec:dns}

\begin{table*}
    \caption{Description of the numerical database. }\label{tab:dns}
    \centering
    \begin{tabular}{c c c c c c c c c c c c}
        \hline\hline
        \\[-0.5em]
        $Re$ & $Ma$ &~~~& $\nu$ & $c_s$ & $N$ &~~~& $\epsilon_f$ & $\eta/\Delta_x$ & $R_\lambda$ & $-\langle p \theta \rangle /  \epsilon_f $ & $\epsilon_d/\epsilon_f$\\ 
        \\[-0.5em]
        \hline
        \\[-0.5em]
        263 & 0.25 &~~~& $3.8 \times 10^{-3}$ & 4.00 & ~128$^3$ &~~~& 0.96 & 1.98 & 21.4 & 0.48 \% & ~0.31 \% \\
        263 & 0.50 &~~~& $3.8 \times 10^{-3}$ & 2.00 & ~128$^3$ &~~~& 0.92 & 2.00 & 21.9 & 0.08 \% & ~2.01 \% \\
        263 & 1.00 &~~~& $3.8 \times 10^{-3}$ & 1.00 & ~128$^3$ &~~~& 1.03 & 1.95 & 20.7 & 1.72 \% & 20.21 \% \\
        263 & 2.00 &~~~& $3.8 \times 10^{-3}$ & 0.50 & ~256$^3$ &~~~& 1.30 & 3.67 & 18.3 & 1.00 \% & 43.27 \% \\
        263 & 4.00 &~~~& $3.8 \times 10^{-3}$ & 0.25 & ~256$^3$ &~~~& 1.27 & 3.67 & 18.6 & 0.59 \% & 50.44 \% \\
        \\
        714 & 0.25 &~~~& $1.4 \times 10^{-3}$ & 4.00 & ~256$^3$ &~~~& 0.70 & 2.02 & 41.2 & 0.48 \% & ~0.26 \% \\
        714 & 0.50 &~~~& $1.4 \times 10^{-3}$ & 2.00 & ~256$^3$ &~~~& 0.68 & 2.04 & 41.7 & 0.09 \% & ~2.12 \% \\
        714 & 1.00 &~~~& $1.4 \times 10^{-3}$ & 1.00 & ~256$^3$ &~~~& 0.72 & 2.01 & 40.6 & 1.97 \% & 23.74 \% \\
        714 & 2.00 &~~~& $1.4 \times 10^{-3}$ & 0.50 & ~512$^3$ &~~~& 1.01 & 3.70 & 34.3 & 1.26 \% & 50.44 \% \\
        714 & 4.00 &~~~& $1.4 \times 10^{-3}$ & 0.25 & ~512$^3$ &~~~& 1.00 & 3.70 & 34.5 & 0.87 \% & 54.07 \% \\
        \\
        1886 & 0.25 &~~~& $5.3 \times 10^{-4}$ & 4.00 & ~512$^3$ &~~~& 0.59 & 2.04 & 72.7 & 0.67 \% & ~0.25 \% \\
        1886 & 0.50 &~~~& $5.3 \times 10^{-4}$ & 2.00 & ~512$^3$ &~~~& 0.59 & 2.04 & 73.0 & 0.20 \% & ~2.17 \% \\ 
        1886 & 1.00 &~~~& $5.3 \times 10^{-4}$ & 1.00 & ~512$^3$ &~~~& 0.64 & 2.00 & 69.9 & 4.06 \% & 25.84 \% \\ 
        1886 & 2.00 &~~~& $5.3 \times 10^{-4}$ & 0.50 & 1024$^3$ &~~~& 0.82 & 3.76 & 61.8 & 3.25 \% & 57.43 \% \\ 
        1886 & 4.00 &~~~& $5.3 \times 10^{-4}$ & 0.25 & 1024$^3$ &~~~& 0.91 & 3.67 & 58.9 & 1.96 \% & 57.62 \% \\ 
        \\
        4166 & 0.25 &~~~& $2.4 \times 10^{-4}$ & 4.00 & 1024$^3$ &~~~& 0.62 & 2.23 & 106.1 & 0.68 \% & 0.13 \% \\
        4166 & 0.50 &~~~& $2.4 \times 10^{-4}$ & 2.00 & 1024$^3$ &~~~& 0.59 & 2.25 & 108.6 & 0.22 \% & 2.37 \% \\
        4166 & 1.00 &~~~& $2.4 \times 10^{-4}$ & 1.00 & 1024$^3$ &~~~& 0.59 & 2.25 & 108.7 & 4.84 \% & 26.10 \% \\
        4166 & 2.00 &~~~& $2.4 \times 10^{-4}$ & 0.50 & 1920$^3$ &~~~& 0.75 & 3.98 & ~96.4 & 4.23 \% & 61.35 \%\\
        4166 & 4.00 &~~~& $2.4 \times 10^{-4}$ & 0.25 & 1920$^3$ &~~~& 0.82 & 3.88 & ~91.7 & 2.60 \% & 61.43 \%\\
        \\[-0.5em]
        \hline
    \end{tabular} 
    \tablefoot{The number of simulation points is noted $N$ and the mesh resolution is denoted $\Delta_x$. The kinetic energy injection rate per unit mass (= viscous dissipation) $\epsilon_f = \langle  \vect{f}\cdot \vect{u}\rangle /\rho_0$ is given in units of $u'^3/L~ (=1)$. The Taylor-scale Reynolds number $R_\lambda = u'^2\sqrt{ 5 / (3\nu \epsilon_f) } $. The Kolmogorov length scale is defined by $\eta = \nu^{3/4} / \epsilon_f^{1/4}$ and is given in units of $\Delta_x = 1/\sqrt[3]{N}$. The pressure dilation $\langle p\theta \rangle$ and the dilatation component of the kinetic energy dissipation rate $\epsilon_d = 4 \langle \mu \theta^2\rangle/3$ are reported in percentage of $\epsilon_f$.}
\end{table*}

\begin{figure*}
    \centering
    \includegraphics[width=0.8\linewidth]{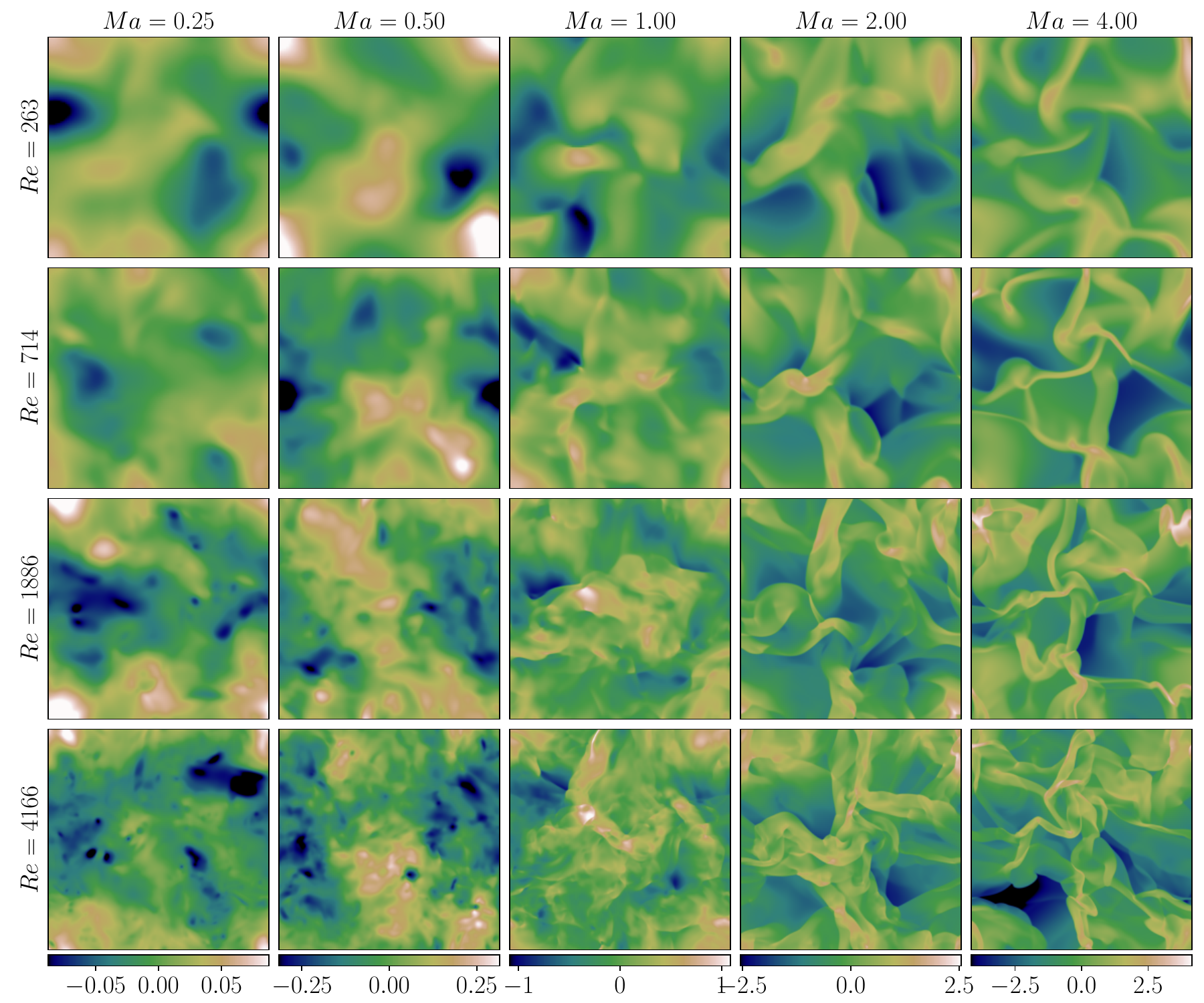}
    \caption{Two-dimensional slices of the density field $s = \ln\rho$ for increasing $Ma$ (from left to right) and increasing $Re$ (from top to bottom). The colorbar spans $\pm 3\sigma_s$.} \label{fig:density_visu}
\end{figure*}

We performed numerical simulations of statistically stationary homogeneous isotropic turbulence. The continuity equation together with the Navier-Stokes equation are solved on a three-dimensional Cartesian mesh, viz
\begin{subequations}
\begin{eqnarray}
    &&\partial_t \rho + \vect{\nabla} \cdot \rho \vect{u}= 0,  \label{eq:continuity}\\
    &&\partial_t \rho \vect{u} + \vect{\nabla} \cdot \rho \vect{u}\vect{u} = -\vect{\nabla} p + \vect{\nabla} \cdot \mathsbf{t} + \vect{f}, \label{eq:ns}
\end{eqnarray}
\end{subequations}
where $\vect{f}$ is a forcing term that is added in order to maintain the turbulent characteristics at statistically steady state. We use the exact same forcing method as in \cite{Federrath2010,Federrath2021}. The force $\vect{f}$ acts at large scales in a band of wavenumbers between $1 \times 2 \pi / L$ and $3 \times 2 \pi / L$, where $L=1$ is the domain width. The forcing spectrum is a parabola which peaks at a wavenumber of $2 \times 2 \pi / L$, and corresponds to a scale of $L/2$. The forcing is composed of a half solenoidal and half compressive mode, termed natural mixture by \citet{Federrath2010}. The reason for this choice is that the natural mixture is believed to be an intermediate case of all possible driving mode occurring in dense molecular clouds which ranges from purely solenoidal to purely compressive \citep{Federrath2010,Gerrard2023}. The code for generating the forcing field $\vect{f}$ is available on GitHub \citep{Federrath2022}. An isothermal equation of state for a perfect gas, $p = \rho c_s^2$ is used to relate the pressure $p$ and the density $\rho$ through the sound speed $c_s$, which is constant. Hence, we do not solve an equation for the internal energy. The viscous stress tensor, $\mathsbf{t}$, is given by 
\begin{equation}
    \mathsbf{t} = 2 \mu \mathsbf{S} - \frac{2}{3} \mu \theta \mathsbf{I} \label{eq:viscous_stress}
\end{equation}
where $\mathsbf{S} = (\vect{\nabla}\vect{u}+\vect{\nabla}\vect{u}^T)/2$ is the strain rate tensor, $\mathsbf{I}$ is the identity matrix and $\theta = \vect{\nabla}\cdot \vect{u}$ is the velocity divergence (or dilatation). Here, we have made the choice of setting a constant kinematic viscosity $\nu = \mu/\rho$. Indeed, it was found that the time step constraint for the simulation was driven by the viscous term, i.e. $\Delta t \sim \nu / \Delta_x^2 = \mu / \rho \Delta_x^2$ ($\Delta_x$ is the spatial resolution). Hence, using a constant dynamic viscosity $\mu$ leads to much smaller time steps compared to using a constant kinematic viscosity, especially at high Mach number where density variations are large. 

For solving Eqs. \eqref{eq:continuity} and \eqref{eq:ns}, we employ an optimised version of the \texttt{FLASH4} code, described in \cite{Federrath2021}. It is a finite volume code which makes use of a positivity-preserving MUSCL-Hancock HLL5R
Riemann scheme \citep{Waagan2011}. The simulation domain is cubic of width $L=1$ with triply periodic boundary conditions. The mean density in the domain is denoted as $\rho_0 = \langle \rho \rangle$ (the angle brackets represent spatial and time average) and was set to 1. 

Our database consists of 20 different simulations where $Ma$ and $Re$ vary independently (see Table \ref{tab:dns}). For this purpose, the forcing amplitude is adjusted so that the velocity dispersion $u' = \langle |\vect{u}|^2\rangle^{1/2}$ was equal to 1 ($\pm 1\%$) in all cases. The Reynolds number $Re = u'L/\nu \equiv 1/\nu$ and Mach number $Ma = u'/c_s \equiv 1/c_s$ were then varied independently by changing the kinematic viscosity $\nu$ and sound speed $c_s$, respectively. The database covers the range $263 \leq Re \leq 4166$ and $0.25 \leq Ma \leq 4.00$, as reported in Table \ref{tab:dns}. For each simulation, after a transient of about $4L/u'$, a statistically steady state is reached. Simulations are then run for 16 additional $L/u'$. All statistics presented hereafter are gathered during the steady state period. For all quantities, the statistical error is estimated as the standard deviation of the mean, divided by the square root of the number of samples. As shown later, statistical uncertainty is affecting only the far tails of the PDFs, and hence only marginally the variance of the distributions.

Care has been taken to ensure an appropriate resolution of the smallest scales of the flow. For incompressible flows, a resolution $\eta = 2\Delta_x$, where $\eta$ is the Kolmogorov length scale, is generally assumed to be sufficient for the velocity gradients to be accurately resolved \citep{Yeung2018}. Hence, for sub- (though slightly compressible) and transonic cases ($Ma \leq 1$), we have adjusted the kinematic viscosity to respect this criterion. 

For the supersonic cases, a good estimate of the adequacy of the numerical resolution is the value of the pressure dilatation term $\langle p \theta \rangle$. Indeed, as per \citet{Pan2019a}, this term represents the reversible transfer between the kinetic energy and internal energy reservoirs. It should therefore be zero if the flow is isothermal and at statistically steady state. When using Riemann solvers without explicitly accounting for a viscous term (an ILES of the Euler equation), \citet{Pan2019a} observed that the statistics of the post-computed dilatation are ill-estimated. They attributed this effect to the numerical schemes which artificially enforces mass conservation at the sacrifice on some errors on $\theta$. They also showed that this bias is mitigated when considering an explicit viscous term as done by \cite{Scannapieco2024} and in the present work. Here, in order to keep the pressure dilatation term $\langle p \theta \rangle$ of the order of a few percent of the energy injection rate, it was necessary to double the mesh resolution for $Ma=2$ and $4$, yielding $\eta \approx 4\Delta_x$. 

Table \ref{tab:dns} summarises some key quantities of the numerical database. $N$ is the total number of points used in the simulation. The energy injection $\epsilon_f = \langle \vect{f}\cdot \vect{u}\rangle /\rho_0$ equals the kinetic energy dissipation rate since the flow is at steady state. The Kolmogorov length scale is given by $\eta = \nu^{3/4}/\epsilon_f^{1/4}$ and is given in units of $\Delta_x = L/\sqrt[3]{N}$. The Taylor-scale Reynolds number $R_\lambda = u'^2\sqrt{ 5 / (3\nu \epsilon_f) } $ varies between about 20 and 110. The pressure dilatation $\langle p \theta \rangle$ is given in units of $\epsilon_f$. The latter appears to be negative and thus represents a loss of kinetic energy. Its maximum value is about 5\% of $\epsilon_f$, although generally much smaller, which confirms that the resolution is adequate.

Table \ref{tab:dns} indicates that when $Ma$ increases, the amount of energy injected into the system increases in order to reach the same $u'$. This results in systematically smaller values of $R_\lambda$ between sub- and supersonic conditions at the $Re$. To some extent, this is due to an increasing contribution of the dilatational component of the dissipation rate which writes $\epsilon_d = 4 \langle \mu \theta^2 \rangle /3$ \citep{Jagannathan2016,John2021}. Table \ref{tab:dns} reveals that $\epsilon_d$ is negligible for $Ma \leq 0.5$ but increases up to 25\% of $\epsilon_f$ at $Ma=1$ and roughly 60\% (if not more) of $\epsilon_f$ for $Ma \geq 2$. We note also that the ratio $\epsilon_d / \epsilon_f$ increases substantially up to $Ma=2$ where it seems to saturate at a value of roughly 60\%. In the trans- and supersonic regimes, there is a slight effect of $Re$ which manifests as a small increase of the ratio $\epsilon_d / \epsilon_f$ when $Re$ becomes larger. We note also that the amount of energy injected $\epsilon_f$ first decreases with $Re$ before reaching an asymptotic value for $Re \geq 1886$, i.e. $R_\lambda \geq 70$. This reflects the evolution of the dissipation constant $C_\epsilon = \epsilon_f L_i /u'^3$, where $L_i$ is the integral length scale. Although it is not the scope of the present work, note that the finite value of $\epsilon_f$ when $\nu \to 0$ is known as the dissipative anomaly. For more details, the reader is referred to the work by \citet{John2021} who discussed the plausible existence of a dissipative anomaly for compressible flows.

\section{Results} \label{sec:results}

\subsection{PDFs of density, velocity and their derivatives} \label{sec:results_pdf}

We start by investigating the effect of $Re$ and $Ma$ on the density fluctuations. The two-dimensional slices of $s = \ln \rho / \rho_0 \equiv \ln \rho$ are displayed in Fig. \ref{fig:density_visu}. The first obvious observation is the increase of the density fluctuations when $Ma$ increases. We further note a clear change of the density structure between sub- and supersonic conditions. Indeed, while the density field is smooth and looks rather similar to a passive scalar field for low Mach number, it becomes filamentary and reveals some zones of abrupt variations (fronts or shocks) for supersonic conditions. The effect of increasing Reynolds number is also discernible, which manifests by an increase of the density fluctuations at small scales. This reflects the smoothing effect due to viscosity which increases with viscosity. 

\begin{figure}
\centering
    \includegraphics[width=0.9\linewidth]{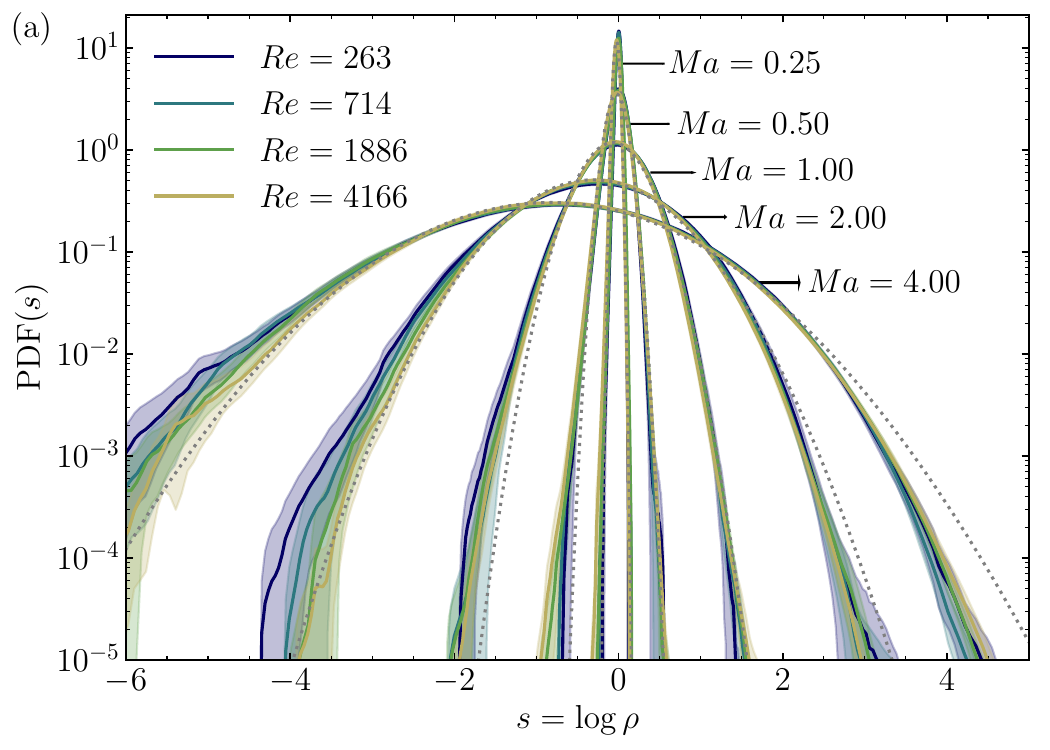}
    \includegraphics[width=0.9\linewidth]{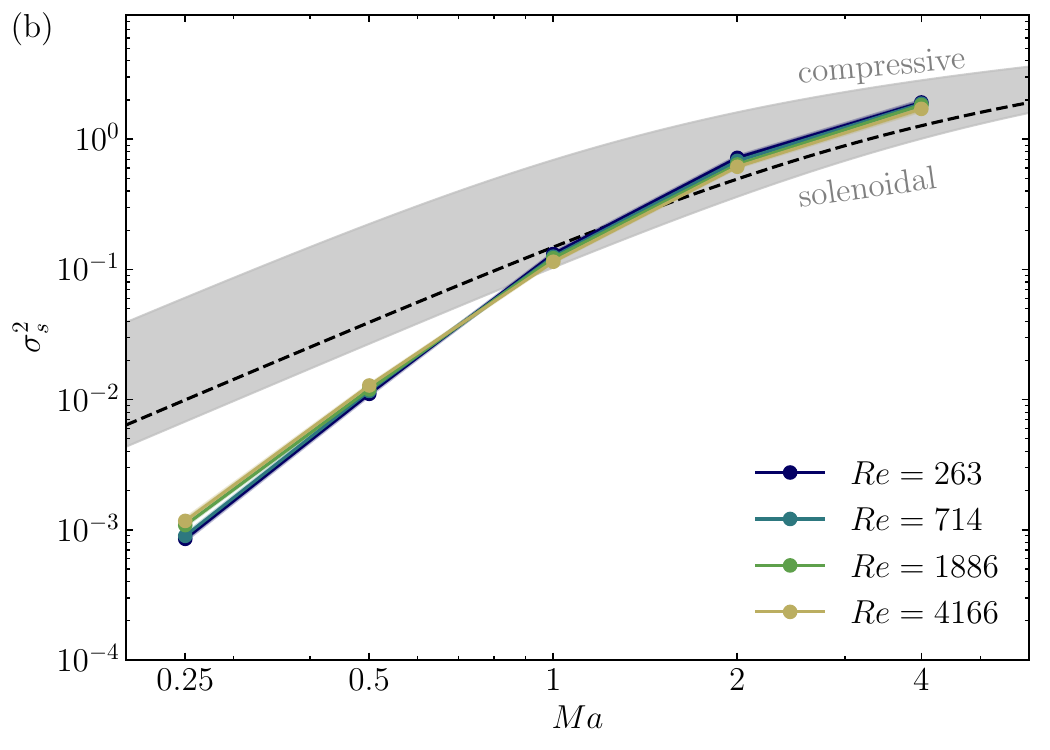}
    \caption{(a) Probability-density-function of $s = \ln \rho$ for varying Mach and Reynolds numbers. Colours from blue to red correspond to increasing Reynolds numbers as shown in the legend. The grey dashed lines are the associated Gaussian distributions (Eq. \ref{eq:log_normal_s}). (b) Evolution of the variance $\sigma_s^2$ with respect to $Ma$. The prediction (Eq. \ref{eq:sigmas_prediction}) is also plotted for different type of forcing (compressive, solenoidal, mixed). The shaded regions represent the statistical uncertainty which affects only the far tails of the PDFs (a) and thus marginally the variance $\sigma_s^2$ (b).}  \label{fig:PDF_rho}
\end{figure}

Quantitative insights can be provided by computing the volume weighted probability-density-functions (PDF) of $s = \ln \rho / \rho_0 \equiv \ln \rho$ which are shown in Fig. \ref{fig:PDF_rho}(a). It is very common \citep{VazquezSemadeni1994,Passot1998,Kritsuk2007,Federrath2008,Federrath2010,Scannapieco2024} to assume that the fluctuations of $\rho$ are log-normal. This prediction arises naturally by assuming a random multiplicative process and the application of the central limit theorem for the density evolution. \citet{Hopkins2013} provides physical arguments showing that density fluctuations may not be log-normal due to intermittency of the field variable. A recent theoretical analysis of \citet{Scannapieco2024} suggests that density fluctuations might follow a stochastic differential process with time-correlated noise. 

Log-normal fluctuations for $\rho$ lead to a Gaussian distribution for $s$, i.e.
\begin{eqnarray}
    {\rm PDF}(s) = \frac{1}{\sqrt{2\pi \sigma_s^2}} \exp \left(-\frac{1}{2} \left(\frac{s-\langle s \rangle}{\sigma_s}\right)^2\right), \label{eq:log_normal_s}
\end{eqnarray}
where $\langle s \rangle$ and $\sigma_s$ denote the mean and standard deviation of $s$, respectively. 

Figure \ref{fig:PDF_rho}(a) indicates that increasing $Ma$ yields larger density fluctuations. The peak of the distribution shifts towards the left as a consequence of mass conservation. Indeed, for a perfect log-normal distribution, mass conservation would have lead to $\langle s \rangle = -\sigma_s/2$ \citep{Passot1998,Kritsuk2007,Federrath2010,Rabatin2023}. In contrast, varying the Reynolds number while the Mach number is constant has a rather marginal effect on the PDF of $s$. This is in agreement with previous numerical results using either an explicit viscous term with different kinematic viscosity \citep{Scannapieco2024}, or an implicit numerical viscosity with different resolutions \citep{Pan2019,Federrath2010}. Note however that although very little, there is a faster drop of the PDFs in the low-density tail when the Reynolds number increases.

Comparing the measured PDFs($s$) to Gaussian distribution reveals that the curves are slightly skewed towards the low-density wing. This effect increases with the Mach number. As suggested by \cite{Federrath2010,Scannapieco2024}, such skewed PDFs appear as a consequence of the compressive part of the forcing. 

The evolution of the variance of $s$, i.e. $\sigma_s^2$, with respect to $Ma$ is shown in Fig. \ref{fig:PDF_rho}(b). Assuming a linear relationship between the density standard deviation $\sigma_\rho$ and $Ma$, viz. $\sigma_\rho = bMa$, and a Gaussian distribution for $s$ (i.e. a log-normal distribution for $\rho$), one arrives at \citep{Federrath2008,Federrath2010,Padoan2011,Molina2012}
\begin{eqnarray}
    \sigma_s^2 = \ln\left(1+b^2Ma^2\right) \label{eq:sigmas_prediction}
\end{eqnarray}
where the coefficient $b$ depends on the type of forcing. For instance, $b=1$ for purely compressive forcing, $b=1/3$ for purely solenoidal and $b \approx 0.4$ for the mixed case used in the present work \citep{Federrath2008,Federrath2010}. Fig. \ref{fig:PDF_rho}(b) indicates that the measured values of $\sigma_s^2$ gradually increase with Mach number. We further note that $\sigma_s^2$ tends to approach the prediction Eq. \eqref{eq:sigmas_prediction} at large $Ma$ only \citep{Konstandin2012,Mohapatra2021}.  
  
\begin{figure}
    \includegraphics[width=0.9\linewidth]{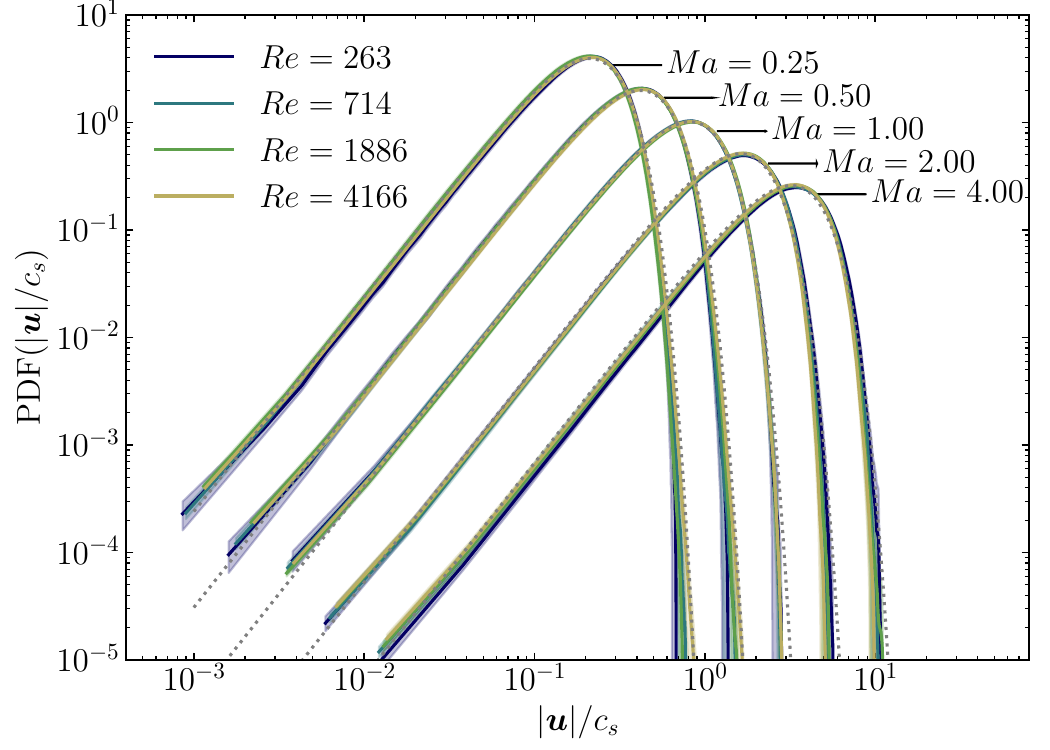}
    \caption{Probability-density-function of the local Mach number $|\vect{u}|/c_s$ for varying $Ma$ and $Re$. The shaded regions represent the statistical uncertainty. The grey dashed lines are the associated Maxwellian distributions Eq. \eqref{eq:log_maxwell}.}\label{fig:PDF_vel}
\end{figure}

We now proceed similarly for the velocity field. For this purpose, we computed the volume weighted probability-density-functions of $|\vect{u}|/c_s$ which can be interpreted as the local Mach number. Recall that in our simulation, the forcing was adjusted so that the variance of $|\vect{u}|$ is constant and independent of the chosen parameters $\nu, c_s$. Assuming that the fluctuations of each component of the velocity vector follow a Gaussian distribution with same standard deviation, one ends up with a Maxwellian distribution for $|\vect{u}|/c_s$ \citep{Rabatin2023}. The latter is given by
\begin{eqnarray}
    {\rm PDF}\left(\left|\vect{u}\right|/c_s\right) = \sqrt{\frac{2}{\pi}} \frac{\left(\left|\vect{u}\right|/c_s\right)^2}{Ma_{1D}^{3}} \exp \left(-\frac{1}{2} \frac{\left(\left|\vect{u}\right|/c_s\right)^2}{Ma_{1D}^2} \right) \label{eq:log_maxwell}
\end{eqnarray}
where $Ma_{1D} = Ma/\sqrt{3}$. The measured PDFs of $|\vect{u}|/c_s$ are plotted in Fig. \ref{fig:PDF_vel}. As previously shown for the density field, the PDFs of $|\vect{u}|/c_s$ appear to be independent of $Re$. They only depend on $Ma$. As expected, increasing $Ma$ yields a shift of the different distributions towards the right and their maximum decreases in proportion of $1/c_s$. That simply means that the PDFs of $|\vect{u}|$ and its components $u_x, u_y, u_z$ are independent of $Ma$ and $Re$. The PDFs of $|\vect{u}|$ would have all collapsed to a single curve. The measured PDFs follow rather nicely a Maxwellian prediction, indicating that the components of $|\vect{u}|$ fluctuate according to a Gaussian distribution. 

\begin{figure*}
    \includegraphics[width=\linewidth]{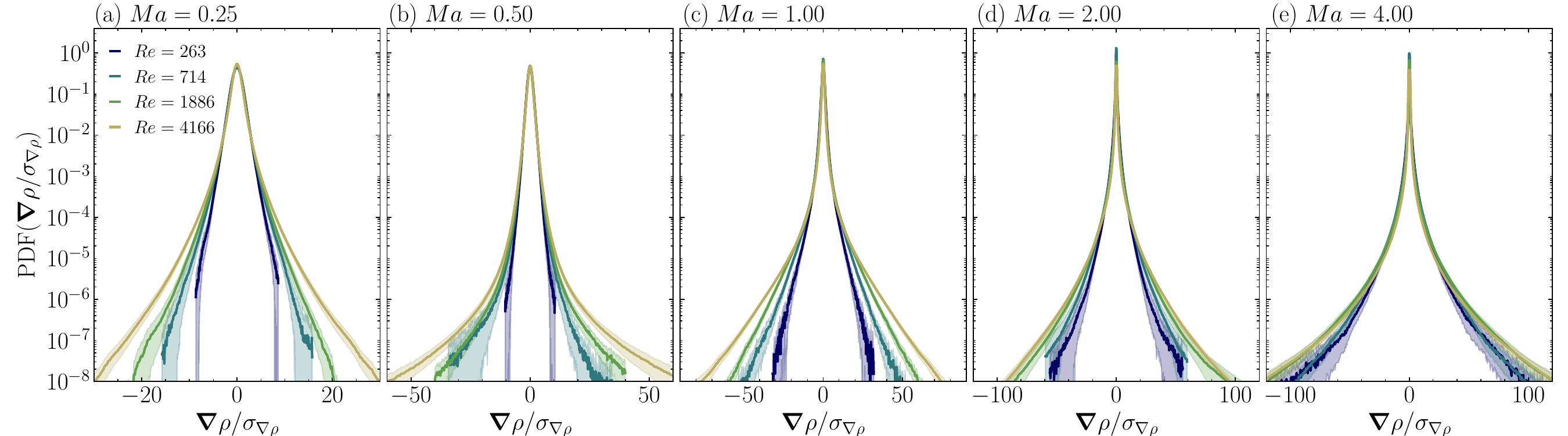}
    \caption{Probability-density-function of the density gradient $\vect{\nabla} \rho$ for varying $Ma$ and $Re$. Colours from blue to red correspond to increasing $Re$ as shown in Fig. \ref{fig:PDF_rho}. The shaded regions represent the statistical uncertainty. Panels (a) to (e) correspond to $Ma=0.25$ to $Ma=4$. } \label{fig:PDF_drho} 
    \includegraphics[width=\linewidth]{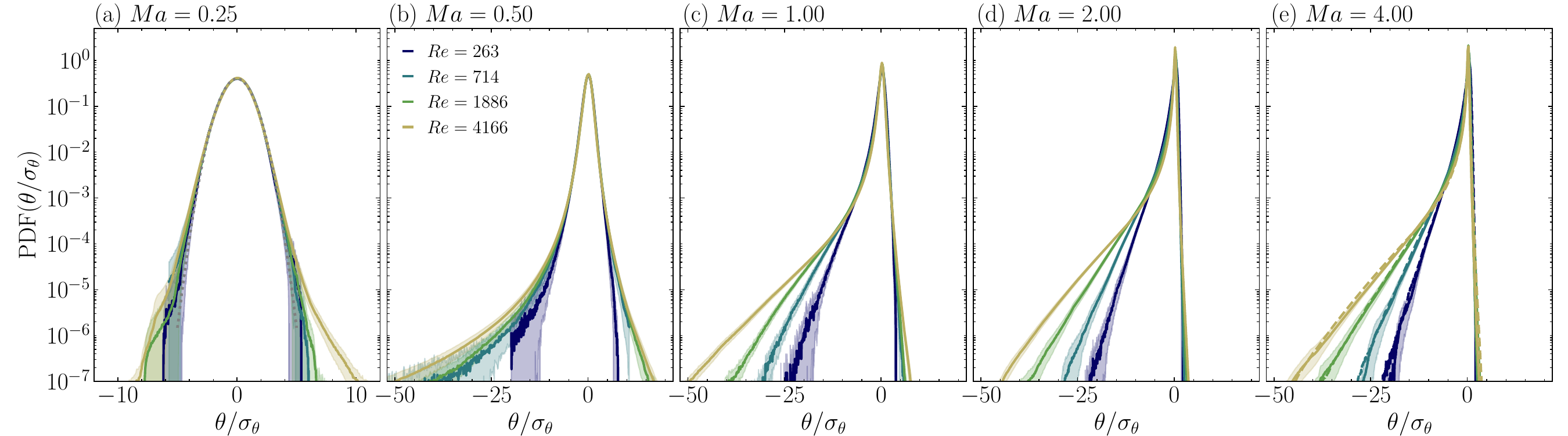}
    \caption{Same as Fig. \ref{fig:PDF_drho} for the velocity divergence $\theta = \vect{\nabla} \cdot \vect{u}$. The dashed lines in panel (e) are the reproduction of the data for $Ma=2$ (panel (d))} \label{fig:PDF_div}
\end{figure*}

The fluctuations of some derivatives of $\rho$ and $\vect{u}$ are also worth quantifying. These appear in particular in the transport equation for $\rho$ (or $s$) together with the transport equation for the kinetic energy explored later in this paper. In Figs. \ref{fig:PDF_drho} and \ref{fig:PDF_div}, we plot the volume-weighted PDFs of $\vect{\nabla} \rho$ and $\theta = \vect{\nabla} \cdot \vect{u}$, respectively. Panels from (a) to (e) are for different $Ma$ while the colour code indicate different $Re$. 

Contrary to $s$ and $|\vect{u}|$, the fluctuations of the density gradient $\vect{\nabla} \rho$ and dilatation $\theta$ depend on both $Ma$ and $Re$. Increasing the Mach and Reynolds numbers yields larger tails in the PDFs for both $\vect{\nabla} \rho$ and $\theta$. This results for sharper density/velocity fronts when compressibility effects (i.e. the Mach number) become larger or viscous effects (i.e. the Reynolds number) get weaker. The measured PDFs deviate very significantly from Gaussian distributions revealing that $\vect{\nabla} \rho$ and $\theta$ are highly intermittent. For instance, at $Ma=4$ and $Re = 4166$, some local fluctuations of $\vect{\nabla} \rho$ can exceed 100 times their standard deviation. Note that in isothermal flows, the pressure $p = \rho c_s^2$ and hence the PDFs portrayed in Fig. \ref{fig:PDF_drho} would be the exact same for $\vect{\nabla} p$. 

The PDFs of dilatation $\theta$ (Fig. \ref{fig:PDF_div}) reveal some skewed distributions that are typical of compressible flows \citep{Passot1998,Jagannathan2016,Schmidt2009,Wang2017,Sakurai2023,Sakurai2024}. At large $Ma$ and $Re$, the left wing of the PDFs, corresponding to compression regions where $\theta <0$, have a much larger extent than the right wing which drops very quickly. Strong compression events are thus more likely than strong expansion events. Despite, the total volume where $\theta <0$ is compensated by the volume formed by $\theta >0$ so that $\langle \theta \rangle = 0$. At the lowest $Ma$, we note that the PDFs($\theta$) are rather symmetric with respect to the $\theta=0$ axis and are almost log-normal. \citet{Pirozzoli2004} and \citet{Sakurai2021}, report similar observation but at a lower Mach number. Comparing the distributions at $Ma = 1, ~2$ and $4$, reveals that the PDFs get narrower when $Ma$ increases from 1 to 2 (Figs. \ref{fig:PDF_div}(c) and (d)).  \citet{Wang2017} made similar observations comparing the PDFs($\theta$) at $Ma=0.8$ and $Ma=1$ (see their Fig. 2). Note that the x-coordinate axis in Figs. \ref{fig:PDF_div} is normalised by $\sigma_\theta$ which increases between $Ma=1$ and $Ma=2$ (this quantity will be discussed in some forthcoming paragraphs). 

Our data further indicate that the PDFs($\theta$) are almost unchanged between $Ma = 2$ and $Ma=4$ (Figs. \ref{fig:PDF_div}(d) and (e)). This is visible in Fig. \ref{fig:PDF_div}(e) where the dashed lines corresponding to $Ma=2$ collapse with the full lines corresponding for $Ma=4$. Speculatively, one could expect that increasing further the Mach number does not have any effects on the PDFs($\theta$). This suggests that for $Ma\geq2$ the distribution of $\theta$ has reached a saturated state, which depends only on the Reynolds number (and probably the type of forcing) but not on the Mach number. 

\begin{figure}
    \centering
        \includegraphics[width=0.9\linewidth]{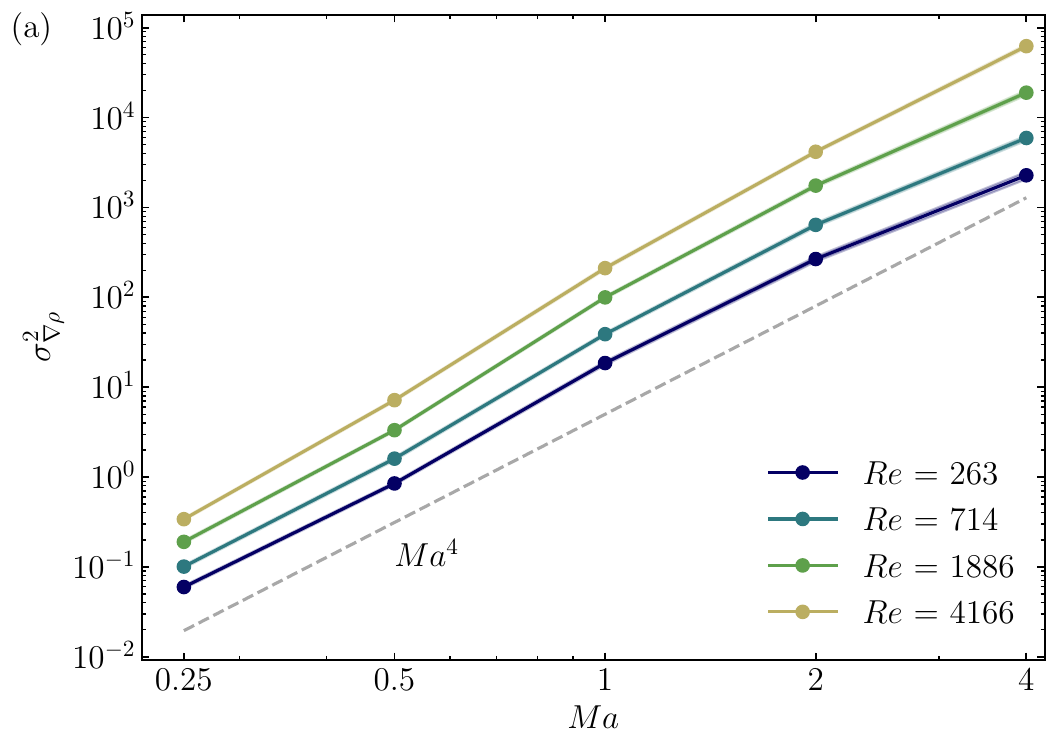}
        \includegraphics[width=0.9\linewidth]{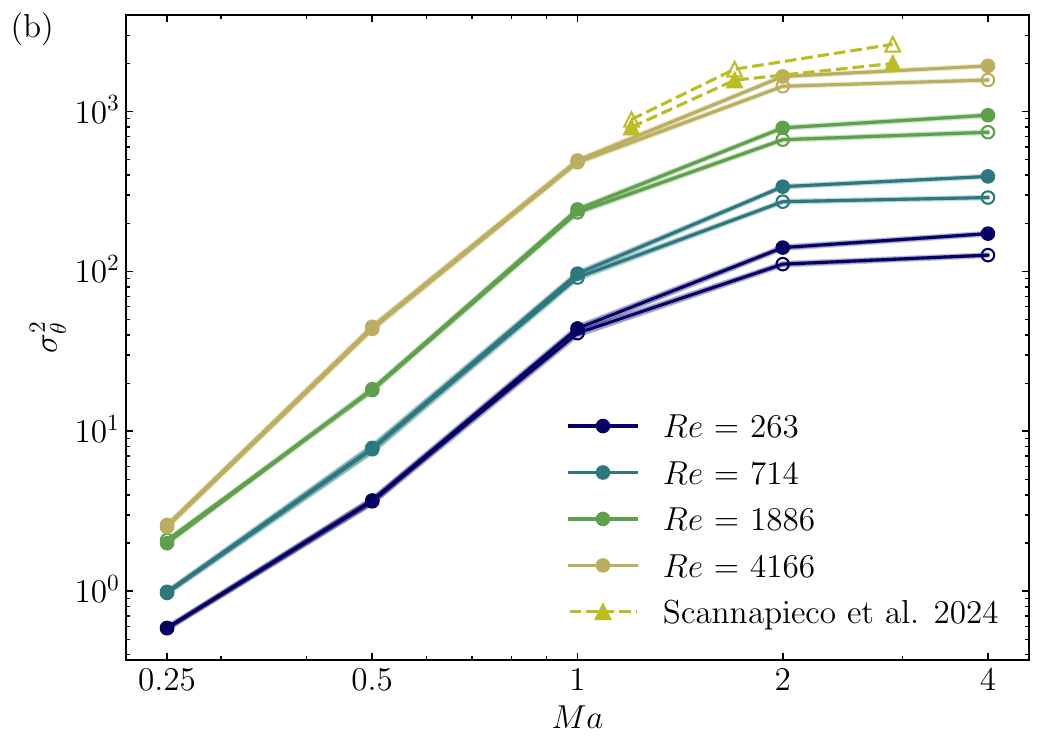}
        \caption{Mach-number dependence of the variance of (a) the density gradient $\vect{\nabla} \rho$ (in units of $\rho_o/L \equiv 1$) and (b) dilatation $\theta$ (in units of $u'/L \equiv 1$) for varying $Re$. In panel (b), the triangle symbols represent the data reported by Ref. \cite{Scannapieco2024} when rescaled in units of $(u'/L)^2$. The closed (open) symbol represent the volume (density) weighted variance.} \label{fig:sigma_div_drho}
    \end{figure}

The variance of $\vect{\nabla} \rho$ and $\theta$, are denoted $\sigma^2_{\nabla \rho}$ and $\sigma^2_{\theta}$, respectively. Their evolution with respect to $Re$ and $Ma$ are reported in Fig. \ref{fig:sigma_div_drho}(a-b). Noticeable is the very strong increase of $\sigma^2_{\nabla \rho}$ when $Ma$ varies from 0.25 to 4 (Fig. \ref{fig:sigma_div_drho}(a)). Our data suggests a power-law variation of the form $\sigma^2_{\nabla \rho} \sim Ma^4$. It also increases with the Reynolds number although with a lower exponent, viz. $\sigma^2_{\nabla \rho} \sim Re^1$. The variance of the dilatation is presented in Fig. \ref{fig:sigma_div_drho}(b). We first note a prompt increase of $\sigma_\theta$ for subsonic (though compressible) conditions before reaching a plateau for $Ma\geq 2$, whose level depends only on the Reynolds number. Here again, we find that the values of $\sigma^2_\theta$ at a given Mach number increase proportionally to $Re$. For comparison, we have plotted the results reported by \citet{Scannapieco2024}. We selected only the cases labelled M1.3$\nu_0$3, M2$\nu_0$3, M3$\nu_0$3 in \cite{Scannapieco2024} as they correspond to roughly the same effective Reynolds number ($Re \sim 3400$ according to our definition). Although quantitatively different, their data are in agreement with our finding that $\sigma^2_{\theta} / (u'/L)^2$ may saturate for $Ma \geq 2$. Since dilatation appears in the vicinity of shocks \citep{Wang2011}, the saturation of $\sigma^2_\theta$ indicates that the shock strength at large $Ma$ is limited by viscous effects. This saturation effect is also observed for the density-weighted variance of $\theta$ (open symbols in Fig. \ref{fig:sigma_div_drho}(b)). Note that since we have used a fluid with a constant kinematic viscosity $\nu$, the density-weighted variance of $\theta$, i.e. $\langle \rho \theta^2 \rangle / \rho_0 = 3 \epsilon_d / (4 \nu)$. Therefore, the saturation of $\langle \rho \theta^2 \rangle$ for $Ma \geq 2$ means that the dilatational component of the kinetic energy dissipation $\epsilon_d$ also saturates to a certain value that uniquely depends on $Re$.

\subsection{Scale-by-scale kinetic energy budgets} \label{sec:results_sbs}

Light is now shed onto the Mach and Reynolds numbers effects on the kinetic energy distribution and kinetic energy transfers. For this purpose, we use a generalised version of the K{\'a}rm{\'a}n-Howarth-Monin (KHM) equation which accounts for the variations of density within the flow. The latter starts from the definition of the turbulent kinetic energy at a given scale proposed by e.g. \citet{Galtier2011,Lai2018,Hellinger2021}, viz.
\begin{eqnarray}
    \langle |\delta \vect{u}|^2_\rho \rangle = \langle (\delta (\rho \vect{u})) \cdot (\delta \vect{u}) \rangle \label{eq:dq2_def}
\end{eqnarray} 
where $\delta (\rho \vect{u}) = (\rho \vect{u})^+ - (\rho \vect{u})^-$ and $\delta \vect{u} = \vect{u}^+ - \vect{u}^-$ represents the increment (the difference) of $\rho \vect{u}$ and $\vect{u}$, respectively, between two points $\vect{x}^+$ and $\vect{x}^-$ arbitrarily separated in space by a distance $\vect{r}$, i.e. $\vect{r} = \vect{x}^+ - \vect{x^-}$. The superscript $+/-$ indicates that the field variable is taken at point $\vect{x}^+/\vect{x}^-$. The brackets in Eq. \eqref{eq:dq2_def} denote a suitable average that depends on the symmetry of the flow. Here, the flow being statistically stationary, homogeneous and isotropic, the average is performed over statistically independent samples, over space and over all possible orientations of the vector $\vect{r}$ using an angular average. By doing this, the quantity $\langle |\delta \vect{u}|^2_\rho \rangle$ depends only on the modulus of the separation vector $r = |\vect{r}|$. It is generally interpreted as the turbulent kinetic energy at a given scale $r$. Note that other definitions for the scale-by-scale turbulent kinetic energy in variable-density flows can be found in the literature \cite[see for instance][]{Ferrand2020,Brahami2020,Hellinger2021a}. It is worth noting that for homogeneous flows, the limit of Eq. \eqref{eq:dq2_def} at large separations yields 
\begin{eqnarray}
    \lim_{r \to \infty} \langle |\delta \vect{u}|^2_\rho \rangle = 2 \langle |\vect{u}|^2_\rho \rangle = 2 \langle \rho |\vect{u}|^2  \rangle \label{eq:dq2_lim},
\end{eqnarray}
where $\langle \rho |\vect{u}|^2  \rangle $ is twice the turbulent kinetic energy. Hence, $\langle |\delta \vect{u}|^2_\rho \rangle$ is not an energy density (such as provided by spectra) but should rather be interpreted as a sort of cumulative kinetic energy distribution of all scales $\leq r$. The transport equation for $\langle |\delta \vect{u}|^2_\rho \rangle$ is known as the K{\'a}rm{\'a}n-Howarth-Monin equation and can formally be written as \citep{Galtier2011,Hellinger2021}:
\begin{equation}
    d_t \langle |\delta \vect{u}|_\rho^2 \rangle = \langle \mathcal{T} \rangle + \langle \mathcal{P} \rangle + \langle \mathcal{V} \rangle + \langle \mathcal{F} \rangle \label{eq:KHM_symbolic}
\end{equation}
The different terms of Eq. \eqref{eq:KHM_symbolic} are explicited below:
\begin{itemize}
    \item $d_t \langle |\delta \vect{u}|_\rho^2 \rangle$ denotes the time variations of $\langle |\delta \vect{u}|_\rho^2 \rangle$. In statistically stationary flows, this term is zero.
    
    \item $\langle \mathcal{T} \rangle$ represents the “transport” of $\langle |\delta \vect{u}|_\rho^2 \rangle$ and stems from the non-linear transport term of the Navier-Stokes equation. In statistically homogeneous flows, the latter can be decomposed into two contributions:
    \begin{align}
      \langle \mathcal{T} \rangle = - \langle \vect{\nabla_r} \cdot (\delta \vect{u}) |\delta \vect{u}|^2_\rho \rangle + \langle \mathcal{R} \rangle. \label{eq:transport} 
    \end{align}   
    The leftmost term on RHS of Eq. \eqref{eq:transport} represents the transfer (or cascade) of kinetic energy among the different scales of the flow. It writes as the divergence in scale-space $\vect{\nabla_r} \cdot$ of the flux  $(\delta \vect{u}) |\delta \vect{u}|^2_\rho$. The term $\langle \mathcal{R} \rangle$ is present only in compressible flows. It represents a source term due to the dilatation and writes
    \begin{eqnarray}
        \langle \mathcal{R} \rangle = \left\langle (\delta \vect{u}) \cdot \left[ (\delta (\rho \vect{u}))(\overline{\delta} \theta) - (\overline{\delta} (\rho \vect{u}))(\delta \theta)\right] \right\rangle,
    \end{eqnarray}
    where $\overline{\delta} \bullet = (\bullet^+ + \bullet^-)/2$ denotes the arithmetic mean of any quantity $\bullet$, between $\vect{x}^+$ and $\vect{x}^-$.
    
    \item $\langle \mathcal{P} \rangle$ represents the effect of pressure on the evolution of the scale-by-scale kinetic energy. In homogeneous compressible flows, it can be decomposed into two contributions:
    \begin{align} 
      \langle \mathcal{P} \rangle = 2 \langle (\delta P) (\delta \theta) \rangle - \langle \mathcal{C}(-\vect{\nabla}p)\rangle  \label{eq:pressure},
    \end{align}  
    The first term on RHS of Eq. \eqref{eq:pressure} is the scale-by-scale contribution of the pressure-dilatation correlation. It corresponds to the scale-by-scale conversion between kinetic energy and internal energy \cite{Aluie2012}. The term $- \langle \mathcal{C}(-\vect{\nabla}P)\rangle$ arises due to variations of the density. The latter was derived by \citet{Hellinger2021} and writes:
    \begin{eqnarray}
        \mathcal{C}(\vect{a}) &:=& (\delta \vect{u}) \cdot (\delta \vect{a}) - (\delta (\rho \vect{u})) \cdot (\delta v\vect{a}) \nonumber \\
        &=& (\rho^+ v^- - 1) \vect{u}^+ \cdot \vect{a}^- + (\rho^- v^+ - 1) \vect{u}^- \cdot \vect{a}^+, \label{eq:Cab}
      \end{eqnarray}
    where $v = 1/\rho$ is the specific volume and $\vect{a}$ denotes any vectorial field quantity. For the pressure term, we set $\vect{a} \equiv -\vect{\nabla}p$. It is easy to show that in constant-density flows, $\rho^\pm v^\mp = 1$ and hence $\mathcal{C}(\vect{a}) = 0$. This term reveals a triple correlation between density, velocity and pressure gradient. It is therefore rather similar to what \cite{Aluie2012} refer to as the baropycnal work, and thus expected to contribute as an additional process of kinetic energy transfer across scales. 

    \item The terms $\langle \mathcal{V}\rangle$ and $\langle \mathcal{F}\rangle$ represent the scale-by-scale contribution of viscous diffusion and forcing, respectively. These terms are here written in the same formulation as \citet{Hellinger2021}, viz.
    \begin{subequations}
        \begin{eqnarray}
            \langle \mathcal{V} \rangle &=&  2 \langle (\delta \vect{u}) \cdot (\delta (\vect{\nabla} \cdot \mathsbf{t})) \rangle - \langle \mathcal{C}(\vect{\nabla} \cdot \mathsbf{t}) \rangle \label{eq:viscous}\\
            \langle \mathcal{F} \rangle &=& 2 \langle (\delta \vect{u}) \cdot (\delta \vect{f}) \rangle - \langle \mathcal{C}(\vect{f}) \rangle  \label{eq:forcing}
        \end{eqnarray}    
    \end{subequations}
    where $\mathsbf{t}$ is the viscous stress tensor defined in Eq. \eqref{eq:viscous_stress} and $\vect{f}$ is the forcing term described in \S \ref{sec:dns}. The expanded version of the viscous term has been derived by \citet{Lai2018}. We keep it here in the compact form given by Eq. \eqref{eq:viscous}.
\end{itemize}

It can be proven that, for statistically homogeneous flows, the limit at large separations of each term of Eq. \eqref{eq:KHM_symbolic} is
\begin{subequations}
    \begin{eqnarray}
        \lim_{r \to \infty} \langle \mathcal{T} \rangle &=& 0, \\ 
        \lim_{r \to \infty} \langle \mathcal{P} \rangle &=& 4 \langle p \theta \rangle, \\
        \lim_{r \to \infty} \langle \mathcal{V} \rangle &=& -4 \epsilon, \\
        \lim_{r \to \infty} \langle \mathcal{F} \rangle &=& 4 \epsilon_f,
    \end{eqnarray}
\end{subequations}
where, $\epsilon = \langle \vect{\nabla}\vect{u} : \mathsbf{t} \rangle$ (the colon operator denotes the double
contraction of second-order tensors) is the kinetic energy dissipation rate. Hence, the one-point kinetic energy budget
\begin{eqnarray}
     d_t \frac{1}{2}\langle |\vect{u}|^2_\rho \rangle = \langle p \theta \rangle - \epsilon + \epsilon_f,
\end{eqnarray}
is recovered from Eq. \eqref{eq:KHM_symbolic} in the limit of large separations ($\rho_0 = 1$ is dropped from the notations). Recall that in statistically stationary flows, both the time derivative and pressure-dilatation terms vanish \citep{Pan2019a} and one ends up with $\epsilon = \epsilon_f$. The kinetic energy dissipation rate $\epsilon = \epsilon_s + \epsilon_d$, where, for homogeneous flows, $\epsilon_s = \langle \mu |\vect{\omega}|^2\rangle$ ($\vect{\omega}$ is the vorticity vector) and $\epsilon_d = 4\langle \mu \theta^2\rangle /3$ are the solenoidal and dilatational components of the dissipation rate, respectively.

\begin{figure}
    \centering
    \includegraphics[width=0.9\linewidth]{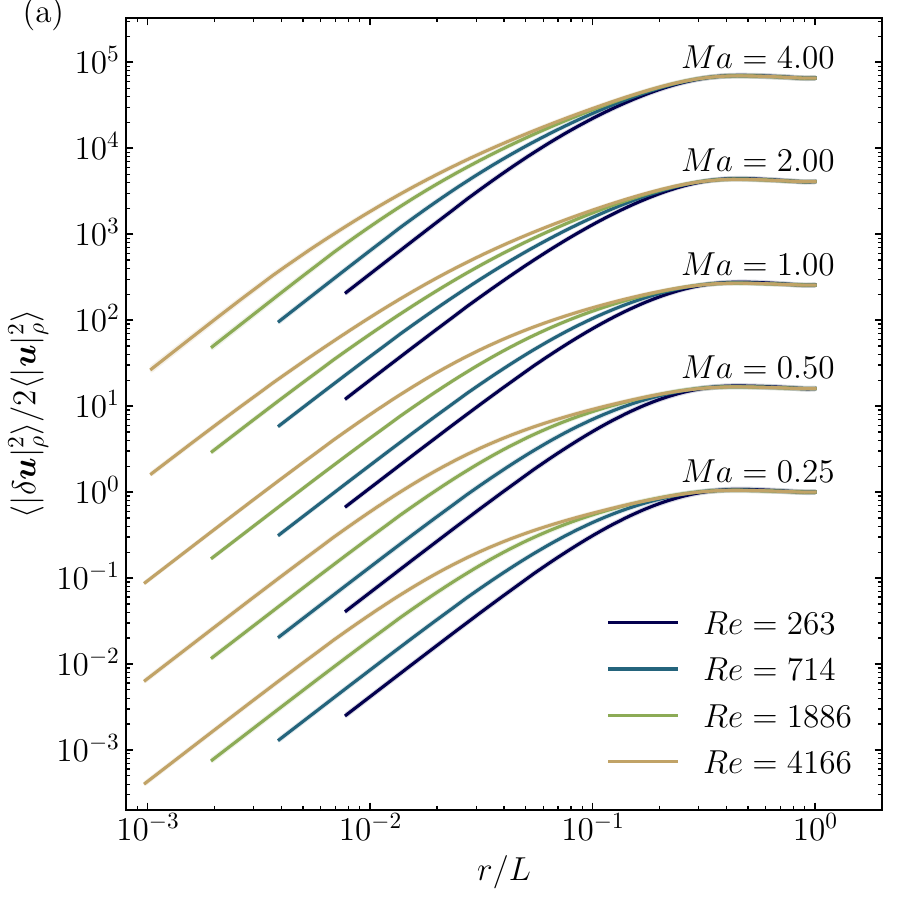} 
    \includegraphics[width=0.9\linewidth]{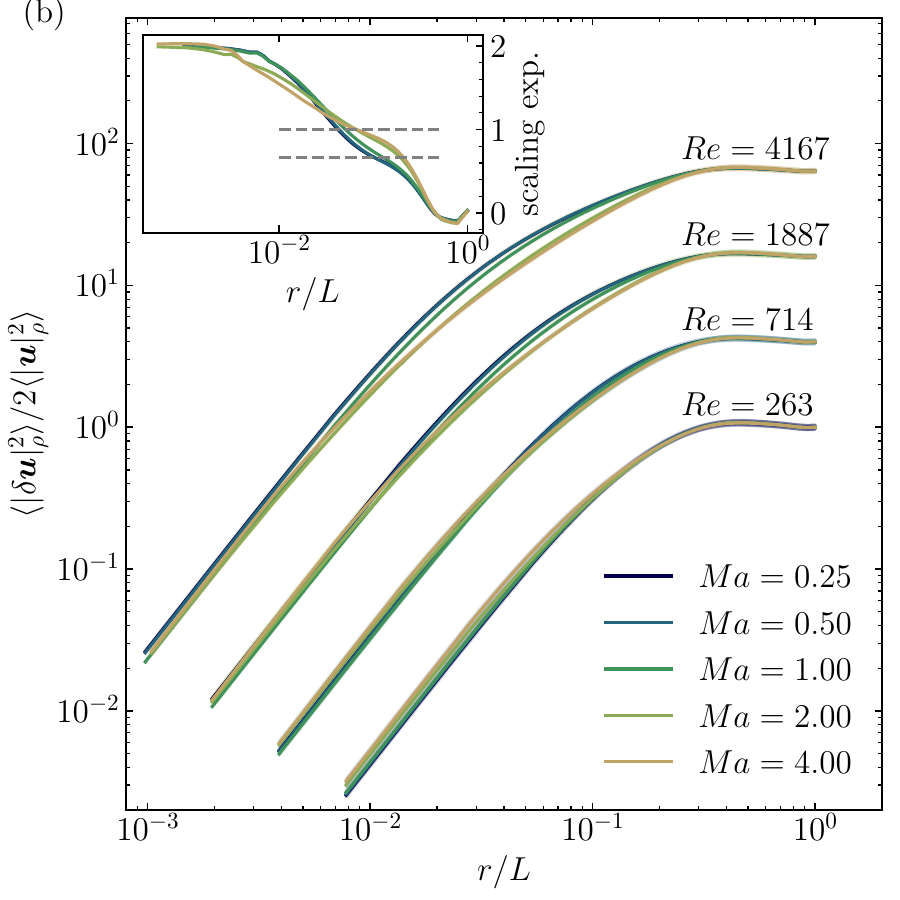}
    \caption{Distribution of kinetic energy $\langle |\delta \boldsymbol{u}|_\rho^2  \rangle/2\langle | \boldsymbol{u}|_\rho^2  \rangle$ as a function of $r/L$. Panel (a) shows the effect of $Re$ at constant $Ma$, while figure (b) does the opposite. The inset in (b) represents the local scaling exponent, Eq. \eqref{eq:scaling_exponent}, at $Re = 4167$. For clarity, each group of curves in (a) and (b) are shifted upwards by a factor 16 and 4, respectively.} \label{fig:dq2}
\end{figure}

The effect of $Re$ and $Ma$ on the scale-by-scale kinetic energy distribution $\langle |\delta \vect{u}|^2_\rho \rangle$ is presented in Fig. \ref{fig:dq2}(a) and (b), respectively. The separation $r$ is normalised using $L$ while $\langle |\delta \vect{u}|^2_\rho \rangle$ is divided by its large-scale asymptotic value, viz. $2\langle |\vect{u}|^2_\rho \rangle$. In Fig. \ref{fig:dq2}(a-b), the different sets of curves corresponding to either constant-$Ma$ or constant-$Re$ are shifted upwards for clarity.

We note that, at constant $Ma$ (Fig. \ref{fig:dq2}(a)),  the different curves expand in the direction of smaller scales when $Re$ increases. This results from the decreasing value of the viscous cut-off scale. A careful examination of Fig. \ref{fig:dq2}(b) reveals that there is no noticeable effect of $Ma$ on $\langle |\delta \vect{u}|^2_\rho \rangle$ at the lowest Reynolds numbers. Its effect becomes substantial only at the largest Reynolds number and in the intermediate range of scales, where one can expect a power-law behaviour of $\langle |\delta \vect{u}|^2_\rho \rangle$ with respect to the scale $r$, i.e. $\langle |\delta \vect{u}|^2_\rho \rangle \sim r^\zeta$. The curves corresponding to the largest Reynolds number in Fig. \ref{fig:dq2}(b) indicate that the scaling exponent is larger for supersonic than subsonic conditions. This can be better quantified by looking at the local scaling exponent defined by:
\begin{eqnarray}
    \zeta(r) = \frac{d \ln \langle |\delta \boldsymbol{u}|_\rho^2 \rangle}{d \ln r}\label{eq:scaling_exponent}
\end{eqnarray}
The latter is portrayed in the inset of Fig. \ref{fig:dq2}(b) for $Re=4167$. Note that the largest Reynolds number studied here is likely too moderate for a clear scaling range to be unambiguously perceived. We observe though that up to $Ma=1$, the scaling exponent in the intermediate range of scales deflects to a value not so far from the Kolmogorov's prediction $\zeta = 2/3$ \citep{Kolmogorov1941}. For $Ma=2$ and $Ma=4$, $\zeta$ deflects to a larger value that seems to comply with the prediction for Burgers turbulence $\zeta = 1$ \citep{Saffman1968,Alam2022}. An analogous observation was reported by \citet{Schmidt2019} using a spectral decomposition of the velocity. Note that towards the smallest scales, the local scaling exponent $\zeta(r) \to 2$ for all values of $Ma$. This agrees with the asymptotic scaling as $r \to 0$ for both Kolmogorov and Burgers turbulence \citep{Kolmogorov1941,Alam2022}.

In summary, in sub- up to transonic conditions, compressibility effects are relatively weak and the scale-by-scale energy distribution is not so different from what one expects for incompressible (Kolmogorov) turbulence. In the supersonic regime, compressibility effects appear to affect mainly the intermediate range of scales, where the scaling exponent is likely to comply with a Burgers kind of turbulence. More studies at higher $Re$ and $Ma$ are likely needed in order to draw firm conclusions about the scaling exponents in the intermediate range of scales. Note that \citet{Federrath2021} revealed that at sufficiently large Reynolds and Mach numbers, both Kolmogorov and Burgers scalings may coexist in two separate range of scales delimited by the so-called sonic scale, i.e. the scale at which the local-in-scale Mach number is 1. From the present data, this double scaling does not appear as a consequence of the moderate Reynolds number.

\begin{figure*}
    \centering
    \includegraphics[width=0.9\linewidth]{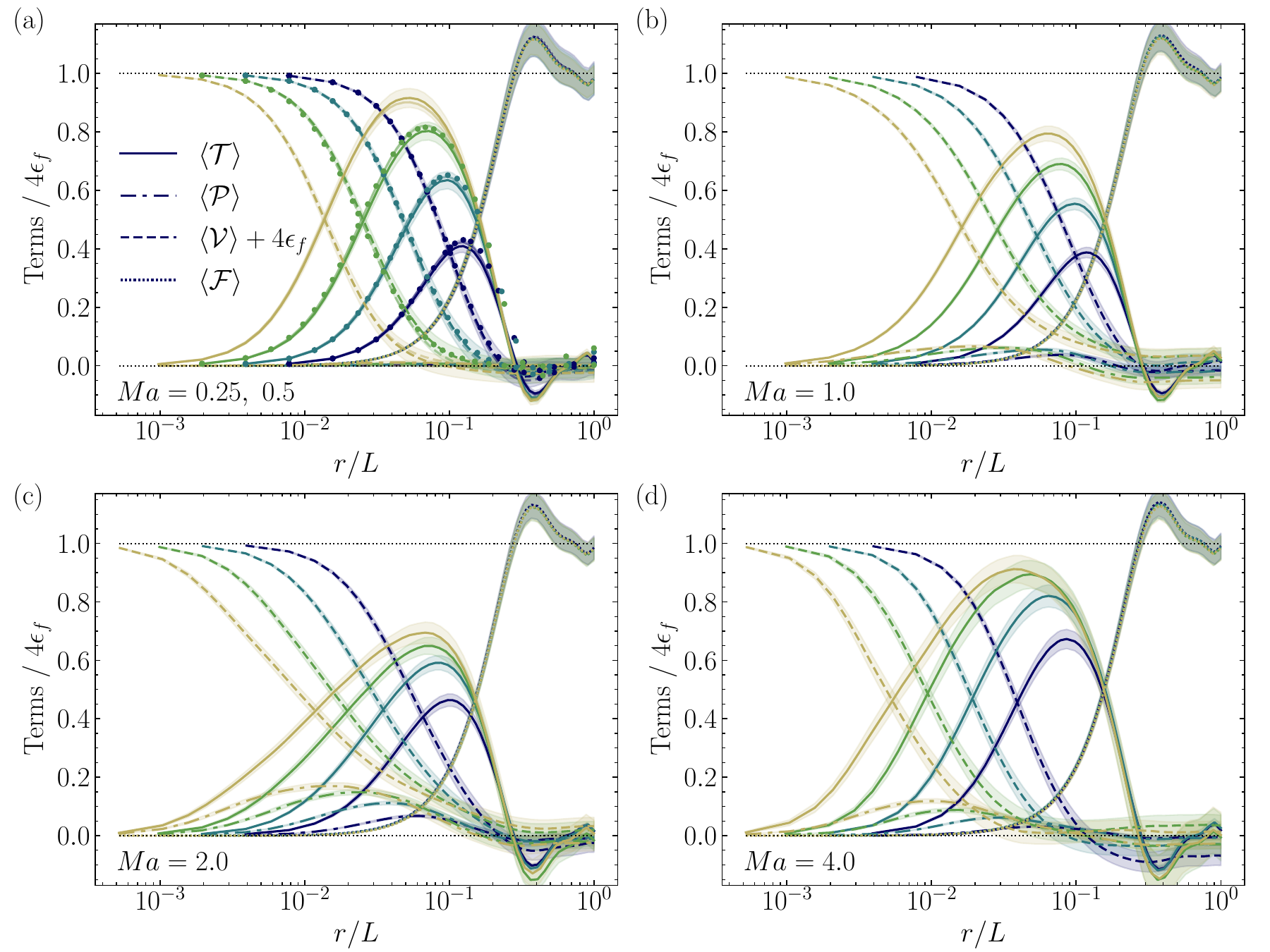}  
    \caption{Scale-by-scale kinetic energy budgets Eq. \eqref{eq:KHM_symbolic} from sub- to supersonic conditions. The different terms of the KHM equation are normalised by $4\epsilon_f$ while the separation is divided by $L\equiv 1$. The color code is same as in Fig. \ref{fig:dq2}(a). The shaded regions represent the statistical error. Panel (a) is for $Ma=0.25$ and $Ma=0.5$, the symbols are the results obtained from an incompressible code. Panel (b), (c) and (d) are for $Ma=1$, $Ma=2$ and $Ma=4$, respectively.} \label{fig:budget}
\end{figure*}

The scale distribution of the different terms of the KHM equation Eq. \eqref{eq:KHM_symbolic} are presented in Fig. \ref{fig:budget}. All terms are divided by $\epsilon_f$. By doing this, the contribution of the forcing term $\langle \mathcal{F}\rangle$ appears to be independent of the conditions investigated. All other terms can then be studied in proportion of an invariant injection of kinetic energy which significantly eases the interpretation of the results. In panel (a), we have combined the results for $Ma=0.25$ (presented with a slight transparency) and $Ma=0.5$. Panels (b) to (d) are for $Ma=1$, $Ma=2$, $Ma=4$, respectively. 

Before we start analysing the KHM equation, it is worth recalling that $\langle |\delta \vect{u}|_\rho^2\rangle$ represents a cumulative energy distribution rather than an energy density. The same applies to the different terms of its transport equation (Eq. \ref{eq:KHM_symbolic}). This is important for interpreting for instance the scale contribution of the transport term $\langle \mathcal{T}\rangle$ and pressure term $\langle \mathcal{P}\rangle$ which are presented in Fig. \ref{fig:budget}(a-d). The latter have a bell shape. They increase when travelling from small to large scales, reach a maximum value and then decrease again to zero at large scales. Since they are cumulative distributions, it means that scales smaller (larger) than the scale at which $\langle \mathcal{T}\rangle$ or $\langle \mathcal{P}\rangle$ is maximum receives (looses) kinetic energy. Hence, such terms are characteristics of the energy cascade or energy conversion between the different scales of the flow. The overall scale-integrated energy transfer/conversion due to the non-linear and pressure terms of the Navier-Stokes equation is zero and hence $\langle \mathcal{T}\rangle\to 0$ and $\langle \mathcal{P}\rangle \to 0$ in the limit of very large separations. The forcing and viscous term do not reveal such a trend but rather asymptote to a non-zero value at large scales which corresponds to the one-point kinetic energy budget given here by $\epsilon_f = \epsilon$. The viscous term $\langle \mathcal{V}\rangle$ in Fig. \ref{fig:budget} is classically presented in the form $\langle \mathcal{V}\rangle + 4\epsilon_f$, and hence $\langle \mathcal{V}\rangle + 4\epsilon_f \to 4\epsilon_f$ when $r\to 0$ and $\langle \mathcal{V}\rangle + 4\epsilon_f \to 0$ when $r \to \infty$. By doing this, the viscous term appears to be confined to small scales in agreement with the theoretical reasoning of \citet{Aluie2013}. Irrespectively of $Re$ and $Ma$, the terms of the KHM equation presented in Fig. \ref{fig:budget}(a-d) indicate that kinetic energy injection is confined to large scales \citep{Aluie2012,Aluie2013}, then transferred/converted to smaller scales by the non-linear transport and pressure terms (when the latter contributes), and is finally dissipated by the viscous term at small scales. 

Looking first to the subsonic cases $Ma=0.25-0.5$ (Fig. \ref{fig:budget}(a)) reveals that the pressure term $\langle \mathcal{P}\rangle$ is negligible. In other words, the cases of $Ma=0.25-0.5$ do not depart significantly from incompressible flows where only the forcing, transport and viscous term contribute to the budget. This is confirmed in Fig. \ref{fig:budget}(a) where the results for $Ma=0.25-0.5$ are compared to their equivalent obtained from a fully incompressible turbulence code \citep{Thiesset2025} using the exact same fluid/forcing parameters. However, it is possible that higher values of $Re$ are needed to perceive compressibility effects at such Mach numbers. The analysis of \citet{Wang2018} at $R_\lambda \approx 250$ indicates that the different terms of the ensemble average coarse-grained kinetic budget are not very sensitive to $Ma$ in the subsonic regime. Figure \ref{fig:budget}(a) indicates that, in the intermediate range of scales, the transport term increases when $Re$ increases. Meanwhile, the viscous term moves towards smaller scales. This behaviour suggests that in the limit of very large $Re$, only the transport term contributes to the budget in the intermediate range, which yields the celebrated Kolmogorov 4/5-law \citep{Nie1999,Antonia2006,Falkovich2010,Danaila2012}, which here can be written as $\langle \mathcal{T} \rangle = 4 \epsilon_f$ for $\eta \ll r \ll L$.

For larger $Ma$ (panel (b) to (d)), the results are rather surprising. For instance, we observe that the contribution of the transport term $\langle \mathcal{T}\rangle$ has a non-monotonic behaviour. Compared to $Ma=0.25$ and $Ma=0.5$, it decreases in amplitude up to $Ma=2$ (Fig. \ref{fig:budget}(c)) before increasing again for $Ma=4$ (Fig. \ref{fig:budget}(d)) to even overtake the incompressible case. A similar, though opposite non-monotonic behaviour applies to the pressure term. The latter increases in amplitude up to $Ma=2$ before decreasing at $Ma=4$. It is likely that increasing further the Mach number results in an even smaller contribution of the pressure term so that $\langle \mathcal{P}\rangle$ may become negligible at very large $Ma$. We also note that $\langle \mathcal{P} \rangle$ has the same shape and same sign as the transport term, although it peaks at smaller scales. Scrutinising the curves for $Ma=1$ to $Ma=4$ (Fig. \ref{fig:budget}(b-d)) indicates that the viscous term tends to contribute at smaller and smaller scales when $Re$ and $Ma$ increases. As a consequence, the total transfer constituted of the non-linear and pressure terms approach $\epsilon_f$ in the limit of large Reynolds numbers. We may thus extend the Kolmogorov 4/5-law to compressible flows by writing $\langle \mathcal{T}\rangle + \langle \mathcal{P} \rangle = 4\epsilon_f$ for $\eta \ll r \ll L$, which can be recast in the same form as Eq. (3.3) of \citet{Falkovich2010}. As already stated, it is plausible that the term $\langle \mathcal{P}\rangle$ vanishes in the limit of large $Ma$ and the generalised Kolmogorov law may reduce to $\langle \mathcal{T}\rangle = 4 \epsilon_f$ for $\eta \ll r \ll L$ when $c_s \to 0$.

\begin{figure*}
    \centering
    \includegraphics[width=0.9\linewidth]{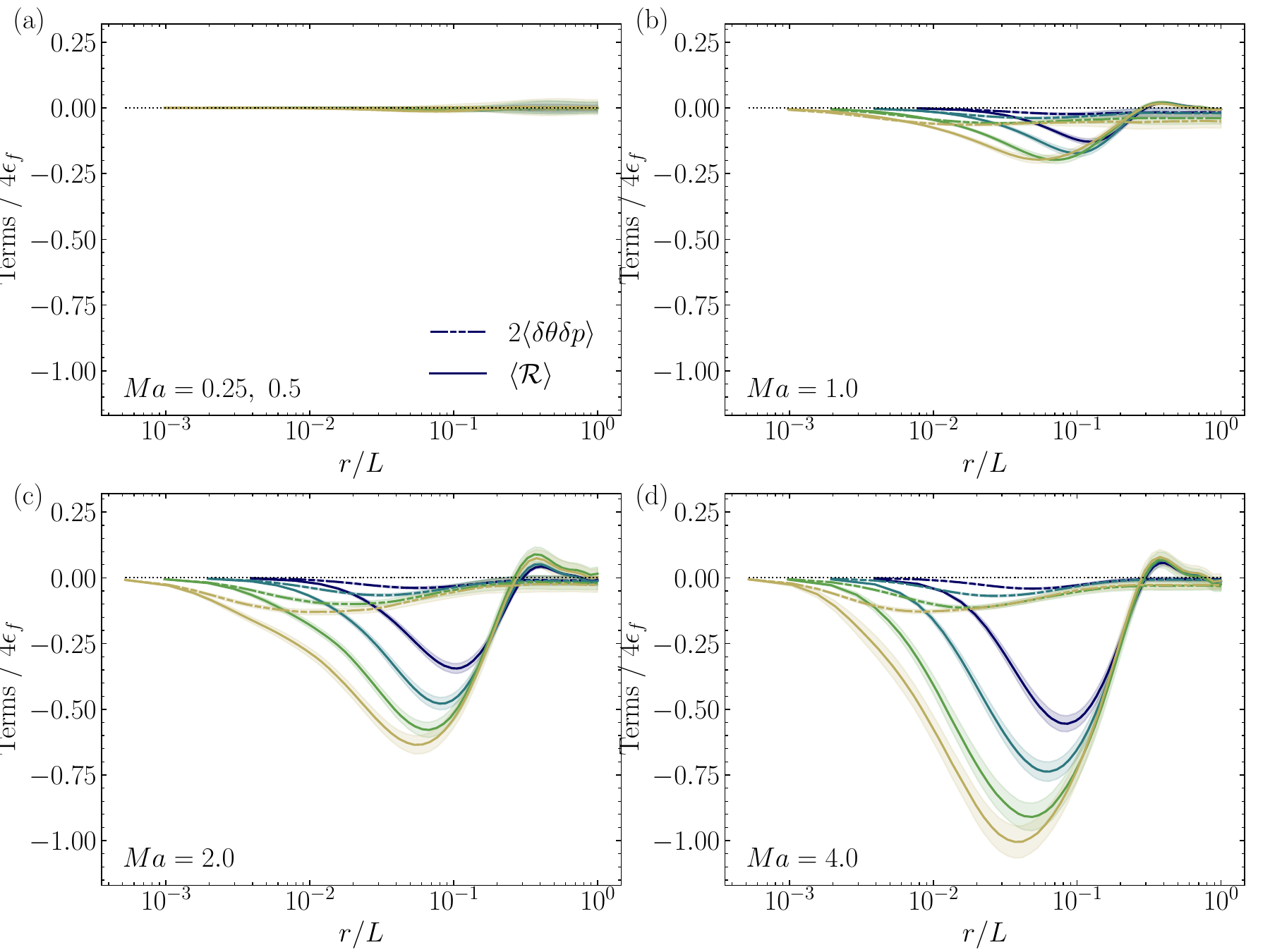} 
    \caption{Scale-by-scale contribution of the terms associated to the dilatation, i.e. $\langle \mathcal{R}\rangle$ and $2 \langle \delta \theta \delta p \rangle$. The color code is same as in Fig. \ref{fig:dq2}(a)} \label{fig:C_budget}
\end{figure*}

We also quantified the scale-by-scale distributions of the terms where dilatation $\theta$ contributes, i.e. the pressure-dilatation term $2\langle \delta \theta \delta p \rangle$ and the source term $\langle  \mathcal{R} \rangle$. Results are presented in Fig. \ref{fig:C_budget}. As observed before, these terms are negligible for $Ma=0.25$ and $Ma=0.5$ and start to contribute significantly only for $Ma \geq 1$. They have the same bell shape as the pressure and non-linear transport term presented before but are mainly negative. This means that dilatation acts as a gain (loss) of kinetic energy for the large (small) scales. Overall, dilatation thus opposes the energy transfer of kinetic energy among the different scales. We further note that the pressure-dilatation term peaks at smaller scales than the source term $\langle \mathcal{R} \rangle$. The latter term $\langle \mathcal{R} \rangle$ appears to increase in amplitude and reaches values close to $4\epsilon_f$ at the largest $Ma-Re$ condition investigated. We do not observe any saturation of this term when $Ma$ and $Re$ increase. In contrast, the two-point pressure-dilatation correlation increases in amplitude up to $Ma=2$ and finally saturates to a certain distribution for larger $Ma$. The saturation of the one-point statistics of dilatation was already identified before in this work. It appears that a similar behaviour applies also for two-point statistics and in particular for the scale-by-scale contribution of the pressure-dilatation term $\langle \delta \theta \delta p \rangle$. Again, this suggests that the dilatation is limited by viscous effects at high $Ma$. 

Overall, our data thus suggest that the kinetic energy exchanges for compressible turbulence operates according to the following scenario. As for incompressible turbulence, part of the energy injected at large scales is transferred to smaller scales through the non-linear transport term. The specificity of compressible flows is that another part of the injected kinetic energy is transferred to smaller scales by the baropycnal work. The pressure-dilatation acts in converting the small-scale kinetic energy into the internal energy reservoir and the opposite conversion occurs at large scales. 

\section{Summary} \label{sec:summary}
We have analysed data from numerical simulations of compressible isothermal turbulence covering different regimes from sub- to trans- to supersonic conditions at different Reynolds numbers. This database extends previous work on the same topic by addressing the case of larger Mach number situations, albeit with some more moderate Reynolds numbers. Care has been taken to the appropriate numerical resolution of the equations by scrutinising the residual of the pressure-dilation term in the one-point kinetic energy budget.

Light is shed on some statistics of the density and velocity fields together with some of their spatial derivatives. We confirm that the statistics of density and local Mach number depend mainly on $Ma$ but not on $Re$. Their spatial derivatives however show a dependence on both $Re$ and $Ma$ which is associated with an increased level of intermittency. {Increased intermittency has significant consequences for the structure and kinematics, as well as the thermodynamics (as the most intermittent structures are believed to be the sites of the most intense dissipation \cite{Falgarone2006,HilyBlant2007,HilyBlant2008}) of the interstellar medium. Indeed, it indicates very local fluctuations of density and velocity and therefore of the kinetic energy dissipation rate with implications for the thermal and chemical evolution of the interstellar medium. }

Analysing the results at larger $Ma$ allowed us to identify that the fluctuations of the dilatation saturate to a certain level that vary with $Re$ but not $Ma$. This indicates that the shock intensity cannot grow indefinitely as it is limited by viscosity. The consequence is that the dilatational component of the kinetic energy dissipation rate $\epsilon_d$ also saturates to a certain proportion of the total kinetic energy dissipation rate $\epsilon$. {Since dilatation plays a crucial role in the dynamic evolution of the density field \citep{Scannapieco2024}, the saturation of its fluctuations may have significant consequences for the density statistics and therefore the star formation rate in astrophysical applications.} 

We then turned our attention to the kinetic energy exchanges among the different scales of the flow. For this purpose we use a generalised version of the KHM-equation, i.e. the transport equation for the scale-by-scale kinetic energy, which accounts for the variations of density and non-zero dilatation. We find that, at constant Reynolds numbers, the effect of Mach number on the kinetic energy distribution is perceived mainly in the intermediate range of scales. We have shown that, for supersonic conditions, the scaling exponent in the "inertial" range approaches Burgers' predictions, while Kolmogorov scaling applies to sub- and transonic conditions. 

The terms of the KHM-equation up to $Ma=0.5$ are identical to their counterpart in incompressible flows. It is not excluded however that, for such values of $Ma$, compressibility effects become perceptible at larger $Re$. For transonic and supersonic conditions, the effect of $Ma$ is rather surprising. Indeed, it is reported that the different terms of the KHM equation have a non-monotonic behaviour. For instance, the maximum transfer of kinetic energy due to the non-linear (pressure) terms of the Navier-Stokes equation appears to decrease (increase) up to $Ma=2$ before increasing (decreasing) again for larger values. The two-point pressure-dilatation correlation saturates to a certain distribution above a certain $Ma$ threshold. Even though this needs to be confirmed by some data at larger $Re$ and $Ma$, our analysis suggests that the generalised 4/5-law proposed by \citet{Falkovich2010}, i.e. $\langle \mathcal{T} \rangle + \langle \mathcal{P} \rangle = \epsilon_f$  may hold in the limit of very large Reynolds numbers. For infinitely large Mach numbers ($c_s \to 0$), our data also indicates that the pressure term is likely to vanish and the 4/5-law may reduce to $\langle \mathcal{T} \rangle = \epsilon_f$. In supersonic situations, the non-linear transport term $\langle \mathcal{T} \rangle$ has a strong contribution due to dilatation, i.e. $\langle \mathcal{R} \rangle$, which represents a loss (gain) of kinetic energy for the small (large) scales. 

{Observational data of the ISM often provides some measures of the velocity structure functions. The KHM is the theoretical framework to interpret these measurements in terms of the underlying physical processes, such as energy injection, transfer, conversion, and dissipation. Our results indicate that $Re$ and $Ma$ significantly influences these processes. We believe that this insight can be useful for better interpreting observational data and refining models of ISM turbulence.}

The present study opens several avenues for future research. As highlighted by \citet{Donzis2020}, the Mach and Reynolds numbers alone do not fully characterise the statistical behaviour of compressible homogeneous turbulence. A more comprehensive description requires the inclusion of an additional parameter: the ratio of solenoidal to dilatational velocity fluctuations. In this work, we have restricted our analysis to the so-called natural mixture of forcing, which consists of an equal proportion of solenoidal and compressive modes. It would therefore be of interest to systematically vary this proportion, thereby modifying the solenoidal-to-dilatational ratio, to assess whether our conclusions hold qualitatively under different forcing conditions. More generally, further insight could be gained by examining scale-local processes after decomposing the velocity field into its solenoidal and dilatational components. Secondly, we have hypothesised that, at sufficiently high $Ma$, dilatational fluctuations are constrained by viscous effects. This conjecture could be tested by quantifying both the width and intensity of shock structures. Thirdly, simulations at higher $Re$ and $Ma$ are required to confirm the emergence of the aforementioned asymptotic relations, analogous to the classical 4/5-law.

\begin{acknowledgements}

{We thank the anonymous referee for their constructive feedback, which has helped us improve the manuscript.} Calculations were performed using the computing resources of CRIANN (Normandy, France), under the project 2018002. We also benefited from resources from the Pawsey Supercomputing Centre (project pawsey0810) in the framework of the National Computational Merit Allocation Scheme and the ANU Merit Allocation Scheme. C.~F.~acknowledges funding provided by the Australian Research Council (Discovery Project DP230102280 and DP250101526), and the Australia-Germany Joint Research Cooperation Scheme (UA-DAAD). We further acknowledge high-performance computing resources provided by the Leibniz Rechenzentrum and the Gauss Centre for Supercomputing (grants~pr32lo, pr48pi and GCS Large-scale project~10391), the Australian National Computational Infrastructure (grant~ek9) and the Pawsey Supercomputing Centre (project~pawsey0810) in the framework of the National Computational Merit Allocation Scheme and the ANU Merit Allocation Scheme. The simulation software, \texttt{FLASH}, was in part developed by the Flash Centre for Computational Science at the University of Chicago and the Department of Physics and Astronomy of the University of Rochester.

\end{acknowledgements}

\section*{Data availability}

{The dataset supporting this study, including both raw and plotting data, comprises approximately 55 TB. These data are available from the authors upon request, subject to availability. Please note that certain raw datasets may no longer be accessible due to storage capacity limitations.}

\bibliographystyle{aa}
\bibliography{aa55653-25.bib}
\end{document}